\documentclass[prb, twocolumn, groupedaddress, showpacs]{revtex4}%
\usepackage{amsfonts}
\usepackage{amsmath}
\usepackage{amssymb}
\usepackage{graphicx}
\usepackage{bm}
\usepackage{subfigure}
\usepackage{color}%
\providecommand{\U}[1]{\protect\rule{.1in}{.1in}}

\begin{document}
\title{ Quantum-oscillation-modulated angular dependence of the positive longitudinal magnetoconductivity and planar Hall effect in Weyl semimetals }
\author{Ming-Xun Deng$^{1}$}
\author{Hou-Jian Duan$^{1}$}
\author{Wei Luo$^{2}$}
\author{W. Y. Deng$^{3}$}
\author{Rui-Qiang Wang$^{1}$}
\email{rqwanggz@163.com}
\author{L. Sheng$^{4,5}$}
\email{shengli@nju.edu.cn}
\affiliation{
$^{1}$ Guangdong Provincial Key Laboratory of Quantum Engineering and Quantum Materials, School of Physics and Telecommunication Engineering, South China Normal University, Guangzhou 510006, China\\
$^{2}$ School of Science, Jiangxi University of Science and Technology,
Ganzhou 341000, China\\
$^{3}$ School of Physics and Optoelectronics, South China University of Technology, Guangzhou, Guangdong 510640, China \\
$^{4}$ National Laboratory of Solid State Microstructures and Department of Physics, Nanjing University, Nanjing 210093, China\\
$^{5}$ Collaborative Innovation Center of Advanced Microstructures, Nanjing
University, Nanjing 210093, China}
\date{\today }

\begin{abstract}
We study the positive longitudinal magnetoconductivity (LMC) and planar Hall effect
in Weyl semimetals, following
a recent-developed theory by integrating the Landau quantization with
Boltzmann equation. It is found that, in the weak magnetic field
regime, the LMC and planar Hall conductivity (PHC) obey $\cos^{6}\theta$ and $\cos^{5}\theta\sin
\theta$ dependence on the
angle $\theta$ between the magnetic and electric
fields. For higher magnetic fields, the LMC and PHC cross
over to $\cos^{2}\theta$ and $\cos\theta\sin\theta$ dependence, respectively.
Interestingly, the PHC could exhibit quantum oscillations
with varying $\theta$, due to the periodic-in-$1/B$ oscillations of the chiral
chemical potential. When the magnetic and electric fields are noncollinear,
the LMC and
PHC will deviate from the classical
$B$-quadratic dependence, even in the
weak magnetic field regime.

\end{abstract}

\maketitle
\affiliation{$^{1}$ National Laboratory of Solid State Microstructures and Department of
Physics, Nanjing University, Nanjing 210093, China}
\affiliation{$^{2}$ Collaborative Innovation Center of Advanced Microstructures, Nanjing
University, Nanjing 210093, China}
\affiliation{$^{3}$ Jiangsu Key Laboratory for Optoelectronic Detection of Atmosphere and
Ocean, Nanjing University of Information Science and Technology, Nanjing
210044, China}

\section{introduction}

Since the prediction of Weyl quasiparticles in pyrochlore
iridates~\cite{PhysRevB.83.205101,Balents36}, the study of
properties of Weyl semimetals (WSMs) in the field of condensed-matter physics, ranging from the unique electronic structures to fascinating topological transports, has attracted much attention on both theoretical and experimental sides\cite{PhysRevLett.107.186806,PhysRevLett.107.127205,PhysRevB.84.075129,PhysRevB.85.035103,PhysRevLett.111.246603,HOSUR2013857}.
WSMs are gapless in the bulk, yet possess topologically protected boundary states on the surfaces nonorthogonal to the momentum difference between paired Weyl nodes, which exhibit different topological properties from three-dimensional topological insulators~\cite{Bernevig106802}. Weyl nodes, always coming in pairs in the momentum space, are double degenerate band
touching points of the relativistic linear-dispersion
excitations~\cite{PhysRevB.83.205101, Zyuzin115133, Goswami245107}. A pair of Weyl nodes, with chirality quantum number protected by the quantized Berry (or Chern)
flux, play the parts of the source and sink of Berry curvature in the momentum space~\cite{Volovik2003,Nielsen389}. The projections of a pair of bulk Weyl nodes on the surface Brillouin zone are connected by an open
Fermi arc of the surface states~\cite{PhysRevB.83.205101,
Hosur195102,Okugawa235315,Haldane0529,Potter5161,Imura245415,Lu096804}.

%

A Dirac point can split into a pair of Weyl nodes, in the presence of
perturbations breaking either the time-reversal or spatial-inversion symmetry.
The separation of the Weyl nodes can induce peculiar topological properties,
which endow WSMs with multiple interesting physics, such as the positive
longitudinal magnetoconductivity (LMC)~
\cite{Nat.Phys.11.728,Shekhar2015,Xiong413,ncomms10137,ncomms13142,LvPRL096603,PhysRevX.5.031023,ncomms10301,ncomms10735,PRB041104R,FP127203,China657406}%
and giant planar Hall effect~\cite{PhysRevB.96.041110, PRL176804}. These
fascinating magnetotransport phenomena of WSMs are related to the chiral
anomaly~\cite{Nielsen389}, which refers to the violation of the separate number
conservation laws of Weyl fermions of different chiralities. Nonorthogonal
electric and magnetic fields can pump Weyl fermions between Weyl nodes of
opposite chiralities, and create a population imbalance between them. The
relaxation of the chirality population imbalance contributes an extra electric
current to the system, and therefore results in a positive LMC (or negative
magnetoresistance) and giant planar Hall conductivity (PHC). The anomalous LMC and
PHC, as exotic macroscopic quantum phenomena, have been enjoying a surge
of experimental~\cite{Shekhar2015,Xiong413,ncomms10137,ncomms13142,LvPRL096603,PhysRevX.5.031023,ncomms10301,ncomms10735,PRB041104R,FP127203,China657406}
and
theoretical~\cite{PhysRevLett.111.246603,PhysRevB.85.241101,PRB165101,PhysRevB.88.104412,PhysRevLett.113.247203,PhysRevB.91.245157}
research interest.

While the positive LMC, as a manifested effect of the chiral anomaly in WSMs, has been observed experimentally, its measured dependence on the angle $\theta$ between the electric and magnetic
fields turns to be not quite in line with the theoretical predictions\cite{Xiong413}.
%
%
Primitively, the theory based on the classical Boltzmann theory predicted a $\cos^{2}\theta$ dependence, due to the $B$-quadratic dependence of the chiral anomaly contribution to the
conductivity~\cite{PhysRevLett.113.247203,PhysRevB.88.104412,PhysRevB.91.245157,PhysRevLett.117.136602,PRL176804}%
, where $B$ is magnitude of the magnetic field. The experimentally-observed angular dependence~\cite{Xiong413,ncomms10735}, however, appears to be much stronger than the theoretically-predicted $\cos^{2}\theta$. Burkov
connected the angular narrowing phenomenon with the PHC, and argued that the
presence of both the LMC with a characteristic angular dependence and giant
PHC could be served as a smoking gun signature of the chiral
anomaly~\cite{PhysRevB.96.041110}. On the other hand, as discussed in Ref.\cite{DengPRL},
with increasing the magnetic field, the $B$-quadratic-dependent positive LMC will cross over to be $B$-linearly scaled, which implies that the $\cos^{2}\theta$ dependence will be modified for stronger magnetic fields. Meanwhile, the chiral chemical potential exhibits a periodic-in-$1/B$ quantum oscillation behavior~\cite{DengPRL}, and therefore the angular dependence of the positive LMC
would be very complicated, and their description may be
beyond the usual classical theory. It is of importance to understand theoretically how the
quantum oscillations of the chiral chemical
potential influence the PHC, since it is helpful for us to
understand the angular dependence of the magnetoconductivity, and it could
also provide new perspectives to experimentally identify WSMs.

In this paper, we follow the theory developed in Ref.\cite{DengPRL} and
investigate the effect of quantum oscillations of the chiral anomaly on the PHC and angular
dependence of the positive LMC. We find that the LMC $\Delta\sigma_{zz}(B)$ and PHC
$\Delta\sigma_{xz}(B)$, in the weak magnetic field regime, are scaled
with $B^{2}\cos^{6}\theta$ and $B^{2}\cos^{5}\theta\sin\theta$, respectively,
yielding stronger angular dependence than that obtained by Nandy $et$
$al$.\cite{PRL176804}. For higher magnetic fields, i.e., $B\gg E/\upsilon
_{\mathrm{F}}$, the angular dependence of the LMC and PHC recover those given by Nandy $et$
$al$, with $\Delta\sigma_{zz}(B)\propto B^{2}\cos^{2}\theta$ and $\Delta
\sigma_{xz}(B)\propto B^{2}\cos\theta\sin\theta$. If the Fermi level is
slightly way from the Weyl nodes, a step change occurs in the LMC and PHC,
as $\theta$ reaches a critical value. For higher Fermi
energy, the PHC could oscillate with $\theta$, due to the
periodic-in-$1/B$ oscillations in the chiral chemical potential. When the
magnetic and electric fields are noncollinear, the LMC and
PHC will deviate from the classical
$B$-quadratic dependence even in the weak magnetic field regime.

The rest of this paper is organized as follows. In Sec.\ \ref{MHM}, we
introduce the model Hamiltonian and solve the spectrum for a WSM in the
presence of crossed magnetic and electric fields. The properties of the DOSs
and anomalous magnetotransport are analyzed in details in Sec.\ \ref{DOSs} and
Sec.\ \ref{PHE}, respectively. The last section contains a summary.

\section{Hamiltonian and spectrum}

\label{MHM}

\begin{figure}[ptb]
\centering
\includegraphics[width=0.8\linewidth]{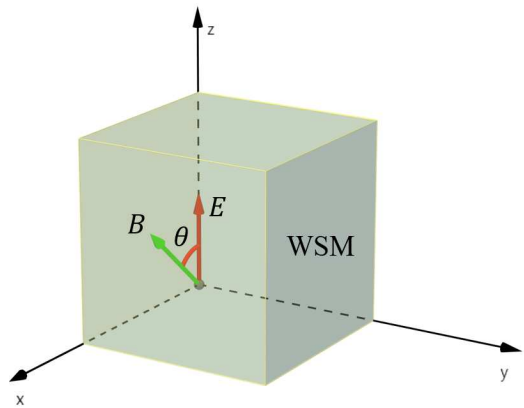} \caption{Schematic
illustration of a WSM subjected to crossed electric and magnetic fields, in
which the electric field is fixed to the $z$-direction and the magnetic field
is confined in the $x$-$z$ plane, with $\theta$ as the included angle between
the electric and magnetic fields. }%
\label{figM}%
\end{figure}

Let us consider a WSM subjected to crossed electric and magnetic fields, which can be described by a low-energy effective Hamiltonian $H\mathcal{=}\sum_{\chi
}\int d^{3}\mathbf{x}\Psi_{\chi}^{\dag}h_{\chi}(\mathbf{x})\Psi_{\chi}$. The
fermion field here is a two-component spinor $\Psi_{\chi}=\exp(i\chi
\mathbf{k}_{c}\cdot\mathbf{x})\psi_{\chi}$, which consists of a slowly varying
part $\psi_{\chi}=(c_{\chi\uparrow},c_{\chi\downarrow})$ and a rapidly
oscillating plane wave with $\chi\mathbf{k}_{c}$ for the momentum locations of
the Weyl nodes and $\chi=\pm$ for chiralities, respectively. The Hamiltonian density for each Weyl node is given by%
\begin{equation}
h_{\chi}(\mathbf{x})=\hbar\upsilon_{\mathrm{F}}\boldsymbol{\sigma}^{\chi}%
\cdot\mathbf{\Pi}+e\mathcal{A}_{0},\label{eq_hamil}%
\end{equation}
where $\upsilon_{\mathrm{F}}$ denotes the Fermi velocity, $\boldsymbol{\sigma
}^{\chi}=\chi(\sigma_{x},\sigma_{y},\sigma_{z})$ with $\sigma_{i=x,y,z}$
as the Pauli matrices for electron spin, and $\mathbf{\Pi
}=-i\mathbf{\nabla}+e\mathbf{A/}\hbar$ stands for the gauge covariant
wave-vector operator modulated by an electromagnetic gauge potential
$\mathcal{A}=(\phi,\mathbf{A})$. The electric and magnetic fields are
connected to $\mathcal{A}$ by the general relations $\mathbf{E}%
=-\mathbf{\nabla}\phi$ and $\mathbf{B}=\mathbf{\nabla}\times\mathbf{A}$.
Specifically, we fix the electric field to the $z$-direction and confine the
magnetic field in the $x$-$z$ plane, i.e., $\mathbf{E}=E\hat{e}_{z}$ and
$\mathbf{B}=B(\sin\theta,0,\cos\theta)$, as illustrated in Fig.\ \ref{figM},
where $\theta$ is the included angle between the electric and magnetic fields.
We choose the Landau gauge $\mathbf{A}=B(x\cos\theta-z\sin
\theta)\hat{e}_{y}$ and $\phi=-Ez$. According to the Landau gauge, we rotate the coordinate system, for convenience, about the $y$
axis, i.e., $y^{\prime}=y$ and
\begin{equation}
\left(
\begin{array}
[c]{c}%
x^{\prime}\\
z^{\prime}%
\end{array}
\right)  =\left(
\begin{array}
[c]{cc}%
\cos\theta & -\sin\theta\\
\sin\theta & \cos\theta
\end{array}
\right)  \left(
\begin{array}
[c]{c}%
x\\
z
\end{array}
\right)  .
\end{equation}
Correspondingly, we also perform a unitary transform on the spinor $\psi
_{\chi}=e^{-i\frac{\theta}{2}\sigma_{y}}\overline{\psi}_{\chi}$. Then the
Hamiltonian, in the rotated coordinate system, can be rewritten as
$H\mathcal{=}\sum_{\chi}\int d^{3}\mathbf{x}^{\prime}\overline{\Psi}_{\chi
}^{\dag}\overline{h}_{\chi}(\mathbf{x}^{\prime})\overline{\Psi}_{\chi}$, with
the Hamiltonian density given by
\begin{equation}
\overline{h}_{\chi}(\mathbf{x}^{\prime})=e^{i(\frac{\theta}{2}\sigma_{y}%
-\chi\mathbf{k}_{c}\cdot\mathbf{x}^{\prime})}h_{\chi}(\mathbf{x}^{\prime
})e^{i(\chi\mathbf{k}_{c}\cdot\mathbf{x}^{\prime}-\frac{\theta}{2}\sigma_{y}%
)}.\label{eq_Hbar}%
\end{equation}
In fact, after the coordinate system rotation, the $z^{\prime}$-axis is
parallel to the magnetic field. Therefore, the vector potential reduces to
$\mathbf{A}^{\prime}=Bx^{\prime}\hat{e}_{y^{\prime}}$ and the scalar
potential becomes $\phi(\mathbf{x}^{\prime})=-(E_{\Vert}z^{\prime}-E_{\bot
}x^{\prime})$, where $E_{\Vert}=E\cos\theta$ and $E_{\bot}=E\sin\theta$.

To solve the energy spectrum, let us start from the Dirac equation,
\begin{equation}
i\hbar\frac{\partial}{\partial t}\overline{\varphi}^{\chi}(\mathbf{x}^{\prime
},t)=\overline{h}_{\chi}(\mathbf{x}^{\prime})\overline{\varphi}^{\chi
}(\mathbf{x}^{\prime},t)\ . \label{eq_dirac}%
\end{equation}
The eigenvalue problem of the crossed electric and magnetic fields in graphene
has been solved analytically by Lukose $et$ $al$.\cite{PhysRevLett.98.116802},
Peres $et$ $al$. \cite{Peres2007} and Krstaji\ifmmode \acute{c}%
\else \'{c}\fi{} $et$ $al$.\cite{PhysRevB.83.075427}. The key step is to find
an appropriate Lorentz boost on the time-space coordinate system and a unitary
transform on the wavefunction. To implement this procedure, we multiply the
both sides of Eq. (\ref{eq_dirac}) by $\sigma_{z}$, and then arrive at the 
covariant time-dependent Dirac equation
\begin{equation}
i\hbar\widetilde{\gamma}^{\mu}(\partial_{\mu}+i\frac{e}{\hbar}\mathcal{A}%
_{\mu}^{\prime})\overline{\varphi}^{\chi}(x^{\mu})=0 \label{eq_codirac}%
\end{equation}
with $x^{0}=\upsilon_{\mathrm{F}}t$ and $x^{1,2,3}=x^{\prime},y^{\prime
},z^{\prime}$, where $\widetilde{\gamma}^{0}=\sigma_{z}$, $\widetilde{\gamma
}^{1,2}=-i\chi\sigma_{y,x}$ and $\widetilde{\gamma}^{3}=\chi\sigma_{0}$. In
order to avoid difficulty in imposing periodic boundary condition along the
$z^{\prime}$ direction\cite{Greenwood1958}, we denote the parallel (along the
magnetic field) component of the electric field by a time-dependent vector
potential $\mathcal{A}_{z^{\prime}}=-E_{\Vert}t$ and then $\mathbf{E}%
=-\partial_{x^{\prime}}\phi(\mathbf{x}^{\prime})-\partial_{t}\mathcal{A}%
_{z^{\prime}}$. Therefore, the Dirac equation is translational invariant along
the $y^{\prime}$ and $z^{\prime}$ directions. The wavefunction thus can take
the form%
\begin{equation}
\overline{\varphi}^{\chi}(x^{\mu})=\frac{1}{\sqrt{L_{y^{\prime}}L_{z^{\prime}%
}}}\sum_{\mathbf{k}}e^{i\mathbf{k}\cdot\mathbf{x}^{\prime}}\overline{\varphi
}_{\mathbf{k}}^{\chi}(x^{\prime},t) \label{eq_wavef}%
\end{equation}
with $\mathbf{k}=(0,k_{y},k_{z})$ measured from the corresponding Weyl node
and $\mathbf{x}^{\prime}=(x^{\prime},y^{\prime},z^{\prime})$. The parallel
electric field enters to $k_{z}$ merely as a parameter by the substitution $k_{z}\rightarrow k_{z}+eE_{\Vert}t/\hbar$. To eliminate the
vertical (perpendicular to the magnetic field) component of the electric
field in Eq. (\ref{eq_codirac}), we apply a Lorentz boost in the direction parallel to the vector
potential
\begin{equation}
\left(
\begin{array}
[c]{c}%
\widetilde{x}^{0}\\
\widetilde{x}^{2}%
\end{array}
\right)  =\left(
\begin{array}
[c]{cc}%
\cosh\vartheta & \sinh\vartheta\\
\sinh\vartheta & \cosh\vartheta
\end{array}
\right)  \left(
\begin{array}
[c]{c}%
x^{0}\\
x^{2}%
\end{array}
\right)
\end{equation}
with $\vartheta=\tanh^{-1}\frac{E_{\bot}}{\upsilon_{\mathrm{F}}B}$, and then
perform a unitary transform on the wavefunction $\overline{\varphi
}_{\mathbf{k}}^{\chi}(x^{\prime},t)=e^{-\chi\frac{\vartheta}{2}\sigma_{y}%
}\widetilde{\varphi}_{\widetilde{\mathbf{k}}}^{\chi}(x^{\prime},\widetilde{t}%
)$. After that we can rewrite Eq. (\ref{eq_codirac}) to be%
\begin{equation}
(\widetilde{\gamma}^{0}\widetilde{\partial}_{0}+\widetilde{\gamma}%
^{3}\widetilde{\partial}_{3}+\widetilde{\gamma}^{+}\widetilde{a}_{\xi
}+\widetilde{\gamma}^{-}\widetilde{a}_{\xi}^{\dag})\widetilde{\varphi
}_{\widetilde{\mathbf{k}}}^{\chi}(x^{\prime},\widetilde{t})=0,
\end{equation}
where $\widetilde{\gamma}^{\pm}=(\widetilde{\gamma}^{1}\pm i\widetilde{\gamma
}^{2})/(\sqrt{2}\widetilde{l}_{B})$ and $\widetilde{l}_{B}=\eta l_{B}$, with
$\eta=(1-\tanh^{2}\vartheta)^{-1/4}$ and $l_{B}=\sqrt{\hbar/eB}$ the magnetic
length. The ladder operators are defined as $\widetilde{a}_{\xi}^{\dag
}=(\widetilde{\xi}-\widetilde{\partial}_{\xi})/\sqrt{2}$ and $\widetilde{a}%
_{\xi}=(\widetilde{\xi}+\widetilde{\partial}_{\xi})/\sqrt{2}$, where
$\widetilde{\xi}=(x^{\prime}-i\widetilde{l}_{B}^{2}\widetilde{\partial}%
_{y})/\widetilde{l}_{B}$.

\begin{figure}[ptb]
\centering
\includegraphics[width=\linewidth]{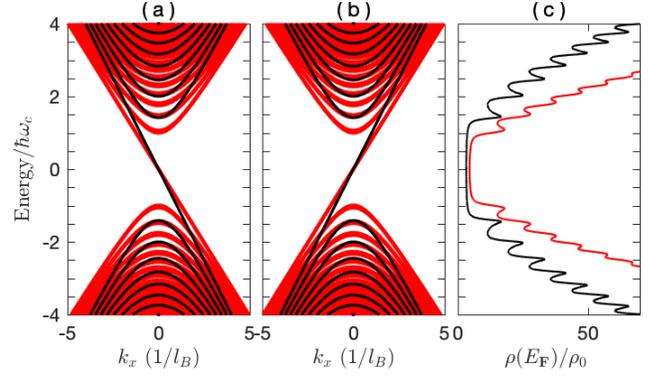} \caption{ The LLs projected to
the $k_{y}=0$ plane for (a) $\chi=+$ and (b) $\chi=-$ Weyl valleys. (c) The
DOSs of the Weyl fermions as functions of the Fermi energy, in which the
characteristic width of the LLs is chosen as $\Gamma=0.05\hbar\omega_{c}$. The
dark (red) curves, for $\theta=0$ ($\pi/5$), represent the electric and
magnetic fields are collinear (noncollinear).}%
\label{figLLs}%
\end{figure}

In the boosted frame, we can derive the time component of the four-momentum as
$\widetilde{\varepsilon}_{n}^{\chi}=\hbar\upsilon_{\mathrm{F}}\Omega_{n,k_{z}%
}^{\chi}$, where
\begin{equation}
\Omega_{n,k_{z}}^{\chi}=s_{n}\sqrt{\frac{2|n|}{\widetilde{l}_{B}^{2}}%
+k_{z}^{2}}-\chi k_{z}\delta_{n,0} \label{eq_tcfm}%
\end{equation}
with $s_{n}\equiv\mathrm{sgn}(n)=\{1,0,-1\}$ for $n\{>,=,<\}0$. Subsequently,
by the inverse Lorentz boost transformation, we can obtain for the spectrum in
the laboratory coordinate system%
\begin{equation}
\varepsilon_{n}^{\chi}(k_{y},k_{z})=\eta^{-2}\hbar\upsilon_{\mathrm{F}}%
\Omega_{n,k_{z}}^{\chi}-\hbar\upsilon_{\mathrm{F}}k_{y}\tanh\vartheta.
\label{spectrum}%
\end{equation}
The eigenstates corresponding to the spectrum are given by
\begin{equation}
\overline{\varphi}_{\mathbf{k}n}^{\chi}(x^{\prime},t)\propto\frac
{e^{-\chi\frac{\vartheta}{2}\sigma_{y}}}{\sqrt{2}}\left(
\begin{array}
[c]{c}%
\chi s_{n}\alpha_{n,+}^{\chi}(k_{z})\phi_{|n|-1}(\xi^{\prime})\\
i\alpha_{n,-}^{\chi}(k_{z})\phi_{|n|}(\xi^{\prime})
\end{array}
\right)  ,
\end{equation}
where $\phi_{n}(\xi)$ are the Landau-gauge orbital wavefunctions, $\xi
^{\prime}=(x^{\prime}+x_{c}^{\prime})/\widetilde{l}_{B}$ and
\begin{equation}
\alpha_{n,\pm}^{\chi}(k_{z})=\sqrt{1\pm\chi k_{z}/\Omega_{n,k_{z}}^{\chi}}.
\end{equation}
When the electric and magnetic fields are collinear, i.e., $\theta=0$, the
Landau levels (LLs) are degenerate, with degeneracy equal to $1/2\pi l_{B}%
^{2}$ per unit cross-section. If the electric and magnetic fields are
noncollinear, as shown by Eq. (\ref{spectrum}), the degeneracy of the LLs will
be lifted and simultaneously, the cyclotron centers given by
\begin{equation}
x_{c}^{\prime}=l_{B}^{2}k_{y}+l_{B}^{2}\Omega_{n,k_{z}}^{\chi}\sinh
\vartheta\label{center}%
\end{equation}
are also renormalized. The spectrums projected to the $k_{y}=0$ plane are
plotted in Figs. \ref{figLLs} (a)-(b), from which we can see that the LLs will
expand to a sequence of Landau bands (LBs) for $\theta\neq0$.

The group velocity for the Weyl fermions equals to the slopes of the spectrum
$\upsilon_{n,\alpha^{\prime}}^{\chi}=\partial\varepsilon_{n}^{\chi}%
(k_{y},k_{z})/\hbar\partial k_{\alpha}$. The velocity along the magnetic
field
\begin{equation}
\upsilon_{n,z^{\prime}}^{\chi}=\eta^{-2}\upsilon_{\mathrm{F}}(s_{n}\frac
{k_{z}}{\sqrt{\frac{2|n|}{\widetilde{l}_{B}^{2}}+k_{z}^{2}}}-\chi\delta
_{n,0}), \label{eq_vnz}%
\end{equation}
scaled with $\eta^{-2}=\sqrt{1-\tanh^{2}\vartheta}$, will decrease with
increasing $\vartheta$, which is also demonstrated in Fig. \ref{figLLs}. As
can be seen from Eq. (\ref{eq_vnz}) and Fig. \ref{figLLs}, the chiral $n=0$ LL
is massless, in which the velocity is $k_{z}$-independent, so that the Weyl
fermions can not be accelerated by the electric field, while the fermions at
the achiral $n\neq0$ LLs, with nonzero effective mass, will be accelerated by
the parallel electric field through $k_{z}+eE_{\Vert}t/\hbar$. Interestingly,
when the magnetic and electric fields are noncollinear, the Weyl fermions, as
indicated by the velocity%
\begin{equation}
\upsilon_{y}=-\upsilon_{\mathrm{F}}\tanh\vartheta=-\frac{E}{B}\sin\theta,
\end{equation}
will drift along the $y$-direction (perpendicular to the electric and magnetic
fields). The drifting velocity, irrelevant to the LL and valley indices, is
inversely (directly) proportional to the magnetic (vertical electric) field.
This result can be understood from the perspective of classical mechanics. In
the plane vertical to the magnetic field, there are two forces, i.e., the electric field and
Lorentz forces, simultaneously acting on a moving Weyl fermion. As a result,
along the direction perpendicular to the electric and magnetic fields, only
those fermions with Lorentz force balanced by the electric field force, e.g.,
$eE_{\bot}=-e\upsilon_{y}B$, can pass through the sample, while others will be
totally filtered out. Therefore, by tuning the directions and relative
magnitudes of the electric and magnetic fields, one can select velocity
[arbitrary directions and magnitudes ($\leqslant\upsilon_{\mathrm{F}}$)] for
the Weyl fermions. This simple mechanism can be used to design velocity
selector for the Weyl fermions.

\begin{figure}[ptb]
\centering
\includegraphics[width=\linewidth]{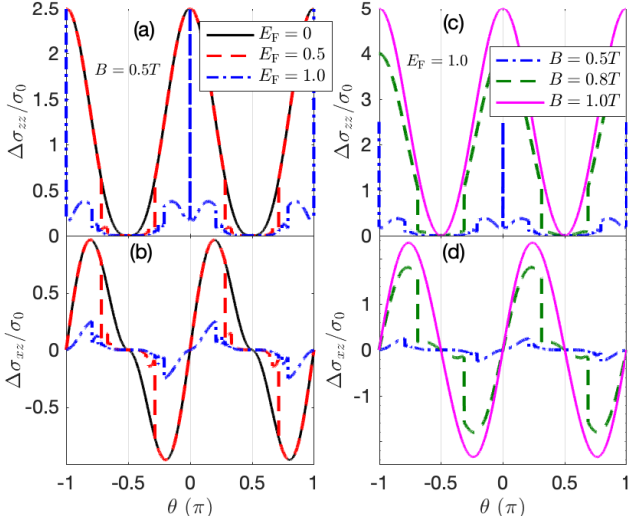} \caption{ The LMC $\Delta
\sigma_{zz}(B)$ (upper panel) and PHC $\Delta\sigma
_{xz}(B)$ (lower panel) as functions of the included angle $\theta$ between
the magnetic and electric fields for several values of (a, b) the normalized
Fermi energy $E_{\mathrm{F}}/\epsilon_{0}$ and (c, d) the magnetic field $B$.
Here and hereafter, for convenience, we choose $\sigma_{0}=\frac{2e^{2}}%
{h}\frac{e[B=1T]\upsilon_{\mathrm{F}}\tau_{\mathrm{inter}}}{h}$ and
$\epsilon_{0}=\upsilon_{\mathrm{F}}\sqrt{\hbar e[B=1T]}$ to be units of the
conductivity and energy, respectively. The magnetic field in (a, b) is fixed
to be $B=0.5T$ and the Fermi energy in (c, d) is set as $E_{\mathrm{F}%
}=\epsilon_{0}$.}%
\label{figLMC}%
\end{figure}

\section{density of states of WSMs subjected to the crossed magnetic and electric fields}

\label{DOSs}

The density of states (DOSs) of a single Weyl valley at the Fermi level can be obtained by the retarded Green's function
\begin{equation}
\rho(E_{\mathrm{F}})=-\frac{1}{\pi}\operatorname{Im}\sum\limits_{n,k_{y}}%
\int\frac{dk_{z}}{2\pi}\frac{1}{E_{\mathrm{F}}+i\Gamma-\varepsilon_{n}^{\chi
}(k_{y},k_{z})}, \label{eq_Gr}%
\end{equation}
where $\Gamma$ characterizes the width of the LLs. The two Weyl valleys have the identical DOSs, whose numerical results are displayed in Fig. \ref{figLLs}(c). For $\Gamma\rightarrow0$, by defining
\begin{equation}
\lambda_{n}(\epsilon)=s_{n}^{2}\sqrt{\epsilon^{2}-2|n|(\eta^{-3}\hbar
\omega_{c})^{2}}+\epsilon\delta_{n,0},
\end{equation}
the DOSs can be derived to be
\begin{equation}
\rho(E_{\mathrm{F}})=\eta^{2}\rho_{0}\left[  2\sum\limits_{n=0}^{n_{c}}%
\frac{|\lambda_{n}(E_{\mathrm{F},+})-\lambda_{n}(E_{\mathrm{F},-})|}{eE_{\bot
}L_{x}}-1\right]  , \label{eq_DOS}%
\end{equation}
where $E_{\mathrm{F},\pm}=E_{\mathrm{F}}\pm eE_{\bot}L_{x^{\prime}}/2$ and
$\rho_{0}=N_{\phi}/h\upsilon_{\mathrm{F}}$, with $N_{\phi}=BS/\phi_{0}$ being
the
degeneracy of the LLs. Here, $S=L_{x^{\prime}}L_{y^{\prime}}$ is the
area of the
cross section perpendicular to the magnetic field, and $\phi_{0}=h/e$ is the
flux quantum. For a relative weak electric field, we can further simplify the
DOSs to be $\rho(E_{\mathrm{F}})=\eta^{2}\rho_{0}\varTheta$, where
\begin{equation}
\varTheta=2\sum\limits_{n=0}^{n_{c}}\frac{1}{\sqrt{1-2|n|(\frac{\eta^{-3}%
\hbar\omega_{c}}{E_{\mathrm{F}}})^{2}}}-1 \label{eq_vartheta}%
\end{equation}
with $n_{c}=\mathrm{int}[\frac{1}{2}(\frac{\eta^{3}E_{\mathrm{F}}}{\hbar
\omega_{c}})^{2}]$ as the index of the highest (lowest) LB crossed by the Fermi
level for $E_{\mathrm{F}}>(<)0$. When the magnetic and electric fields are
collinear, Eq. (\ref{eq_DOS}) recovers the result given in Ref.\ \cite{DengPRL},
leading to quantum oscillations in the DOSs. Here, the van Hove
singularities, determined by
\begin{equation}
E_{\mathrm{F}}=(1-\frac{E^{2}}{\upsilon_{\mathrm{F}}^{2}B^{2}}\sin^{2}%
\theta)^{3/4}\sqrt{2|n|}\hbar\omega_{c}, \label{eq_van}%
\end{equation}
are tunable by the included angle and relative magnitudes of the magnetic and
electric fields. As a consequence, the quantum oscillations are electrically
controllable, as shown by Fig.\ \ref{figLLs}(c).

\section{Field and angular dependence of the LMC and PHC}

\label{PHE}

In this section, following the standard Boltzmann approach, we discuss the
transport properties of the WSM subjected to the crossed electric and magnetic
fields. The steady-state Boltzmann equation for the $n$-th LB of valley $\chi$
is given by%
\begin{equation}
\frac{\partial f_{n}^{\chi}}{\partial t}=\frac{\partial f_{n}^{\chi}}{\partial
t}|_{d}+\frac{\partial f_{n}^{\chi}}{\partial t}|_{c}=0, \label{eq_boltz}%
\end{equation}
where $f_{n}^{\chi}$ is the nonequilibrium electron distribution function and
\begin{equation}
\frac{\partial f_{n}^{\chi}}{\partial t}|_{d}=\frac{\partial\mathbf{k}%
}{\partial t}\cdot(\mathbf{\nabla}_{k}\varepsilon_{n}^{\chi})\frac{\partial
f_{n}^{\chi}}{\partial\varepsilon_{n}^{\chi}}%
\end{equation}
is the drifting
of the distribution function induced by the external electric
field. In the relaxation time approximation, the change of the distribution
function due to electron scattering by impurities
can be expressed as
\begin{equation}
\frac{\partial f_{n}^{\chi}}{\partial t}|_{c}=\frac{f_{n}^{\chi}-f_{\chi}%
}{\tau_{\mathrm{intra}}}+\frac{f_{n}^{\chi}-f_{g}}{\tau_{\mathrm{inter}}}%
\end{equation}
with $\tau_{\mathrm{intra}}$ and $\tau_{\mathrm{inter}}$ being
relaxation
times due to electron intravalley and intervalley scattering by impurities, respectively.
$f_{\chi}$ and $f_{g}=(f_{\chi}+f_{-\chi})/2$ represent the local and global
equilibrium electron distribution functions for the system.
Therefore, Eq.\ (\ref{eq_boltz}) can be rewritten as%
\begin{equation}
e\mathbf{E}\cdot\mathbf{\upsilon}_{n}^{\chi}(\mathbf{k})\frac{\partial
f_{n}^{\chi}}{\partial\varepsilon_{n}^{\chi}}=-\frac{f_{n}^{\chi}-f_{\chi}%
}{\tau_{\mathrm{intra}}}-\frac{f_{n}^{\chi}-f_{g}}{\tau_{\mathrm{inter}}}.
\label{eq_Boltz}%
\end{equation}
Within the framework of the linear response, the electron distribution
function takes the general form%
\begin{equation}
f_{n}^{\chi}(\mathbf{k})=f_{0}(\varepsilon_{n}^{\chi})+[-\partial
_{\varepsilon_{n}^{\chi}}f_{0}(\varepsilon_{n}^{\chi})]g_{n}^{\chi}%
(\mathbf{k}), \label{eq_fn}%
\end{equation}
where $f_{0}(\epsilon)=1/[1+e^{\beta(\epsilon-E_{\mathrm{F}})}]$ stands for
the electron equilibrium distribution function, and $g_{n}^{\chi}(\mathbf{k})$
describes the deviation of $f_{n}^{\chi}(\mathbf{k})$ from $f_{0}%
(\varepsilon_{n}^{\chi})$ due to the applied external fields. Substitution
of Eq.\ (\ref{eq_fn}) into Eq.\ (\ref{eq_Boltz}) leads to%
\begin{equation}
e\mathbf{E}\cdot\mathbf{\upsilon}_{n}^{\chi}(\mathbf{k})=-\frac{g_{n}^{\chi
}(\mathbf{k})-\overline{g}_{\chi}}{\tau_{\mathrm{intra}}}-\frac{g_{n}^{\chi
}(\mathbf{k})-\overline{g}}{\tau_{\mathrm{inter}}} \label{eq_gnc}%
\end{equation}
with $\overline{g}_{\chi}=\langle g_{n}^{\chi}(\mathbf{k})\rangle_{\chi}$ and
$\overline{g}=(\overline{g}_{\chi}+\overline{g}_{-\chi})/2$. The average
$\langle\cdots\rangle_{\chi}$ here is defined as
\begin{equation}
\langle\cdots\rangle_{\chi}=\frac{\sum_{n,k_{y}}\int[-\partial_{\varepsilon
_{n}^{\chi}}f_{0}(\varepsilon_{n}^{\chi})](\cdots)dk_{z}}{\sum_{n,k_{y}}%
\int[-\partial_{\varepsilon_{n}^{\chi}}f_{0}(\varepsilon_{n}^{\chi})]dk_{z}},
\end{equation}
in which the summation runs over all electron states at the Fermi level in the
$\chi$ valley.\begin{figure}[ptb]
\centering
\includegraphics[width=\linewidth]{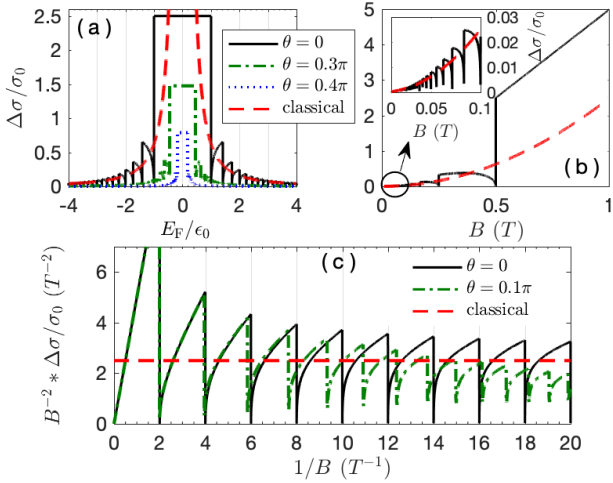} \caption{ Amplitude of the LMC
and PHC $\Delta\sigma(B)$ versus (a) the normalized Fermi
level $E_{\mathrm{F}}/\epsilon_{0}$, (b) the magnetic field $B$ (weak magnetic
field parameter region) and (c) the reciprocal of the magnetic field $1/B$
(strong magnetic field parameter region). The parameters are the same as Fig.
\ref{figLMC}.}%
\label{figPHE}%
\end{figure}

In the absence of the intervalley scattering, the system can not reach the
global equilibrium. As a result, for a given chirality, even at full momentum
relaxation of the electron distribution, there exists a finite electric
current proportional to the chirality imbalance, which predicts an unphysical
diverging electrical conductivity. However, in a real system, the conductivity
can never be infinite. This implies that the intervalley scattering must exist
to relax the system to global equilibrium between the valleys. The emergence
of the LMC requires that the intravalley scattering must be stronger than the
intervalley scattering, i.e., $\tau_{\mathrm{intra}}/\tau_{\mathrm{inter}}%
\ll1$. In this situation, the system would be relaxed to the local equilibrium
first by the intravalley scattering and to the global equilibrium later by the
intervalley scattering. This process would produce a finite chemical
potential difference between the two Weyl valleys, half of which is usually
called as the chiral chemical potential. In the opposite limit, $\tau
_{\mathrm{intra}}/\tau_{\mathrm{inter}}\gg1$, which may occur in Dirac
semimetals, where the energy bands are doubly degenerate, no chiral chemical potential
exists, since the system would be relaxed to the global equilibrium, directly.

In the following, we would consider $\tau_{\mathrm{intra}}/\tau
_{\mathrm{inter}}\ll1$, such that we can safely approximate $g_{n}^{\chi
}(\mathbf{k})=\overline{g}_{\chi}$ in the second term of Eq. (\ref{eq_gnc}),
and obtain for
\begin{equation}
eE_{\Vert}\upsilon_{n,z^{\prime}}^{\chi}(\mathbf{k})=-\frac{g_{n}^{\chi
}(\mathbf{k})-\overline{g}_{\chi}}{\tau_{\mathrm{intra}}}-\frac{\overline
{g}_{\chi}-\overline{g}_{-\chi}}{2\tau_{\mathrm{inter}}}, \label{eq_gn}%
\end{equation}
where $\overline{g}_{\chi}=\chi\Delta\mu$. Here, due to the cyclotron motion,
the Weyl fermions, in the plane perpendicular to the magnetic field, are
localized by the strong magnetic field. Accordingly, we can solve for
\begin{equation}
g_{n}^{\chi}(\mathbf{k})=-eE_{\Vert}\upsilon_{n,z^{\prime}}^{\chi}%
(\mathbf{k})\tau_{\mathrm{intra}}+\chi(1-\frac{\tau_{\mathrm{intra}}}%
{\tau_{\mathrm{inter}}})\Delta\mu. \label{eq_gcn}%
\end{equation}
By averaging the both sides of Eq. (\ref{eq_gcn}), we can determine $\Delta
\mu$ self-consistently to be%
\begin{equation}
\Delta\mu=-\chi eE_{\Vert}\langle\upsilon_{n,z^{\prime}}^{\chi}(\mathbf{k}%
)\rangle_{\chi}\tau_{\mathrm{inter}}=\frac{eE_{\Vert}}{2\pi l_{B}^{2}}%
\frac{\tau_{\mathrm{inter}}}{\mathcal{F}}, \label{eq_muchi}%
\end{equation}
where%
\begin{equation}
\mathcal{F}=\frac{2\pi\hbar}{S}\int_{-\infty}^{\infty}[-\partial_{\epsilon
}f_{0}(\epsilon)]\rho(\epsilon)d\epsilon. \label{eq_F}%
\end{equation}

The electrical current density can be calculated by
\begin{equation}
j_{\alpha}=\frac{e}{2\pi S}%
{\displaystyle\sum\limits_{\chi,n,k_{y}}}
\int\upsilon_{n,\alpha}^{\chi}(\mathbf{k})g_{n}^{\chi}(\mathbf{k}%
)\partial_{\varepsilon_{n}^{\chi}}f_{0}(\varepsilon_{n}^{\chi})dk_{z}
\label{eq_cds}%
\end{equation}
and the conductivity tensor is defined as $\sigma_{\alpha z}=j_{\alpha}/E$. By
substituting Eq. (\ref{eq_gcn}) into Eq. (\ref{eq_cds}), we derive the
conductivity along the electric field to be
\begin{equation}
\sigma_{zz}(B)=[\sigma_{D}+\Delta\sigma(B)]\cos^{2}\theta, \label{eq_szz}%
\end{equation}
where the amplitude is given by
\begin{equation}
\Delta\sigma(B)=\frac{2e^{2}}{h}\frac{(eB)^{2}}{h^{2}}\frac{\tau
_{\mathrm{inter}}-\tau_{\mathrm{intra}}}{\mathcal{F}}%
\end{equation}
and $\sigma_{D}=\frac{2e^{2}\upsilon_{\mathrm{F}}\tau_{\mathrm{intra}}}%
{h}\mathcal{F}_{1}$, with
\begin{equation}
\mathcal{F}_{1}=\frac{1}{2\pi\widetilde{l}_{B}^{2}}\int_{-\infty}^{\infty
}[-\partial_{\epsilon}f_{0}(\epsilon)]\Lambda(\epsilon)d\epsilon
\end{equation}
and%
\begin{equation}
\Lambda(\epsilon)=2%
{\displaystyle\sum\limits_{n=0}^{n_{c}}}
\sqrt{1-2n(\frac{\eta^{-3}\hbar\omega_{c}}{\epsilon})^{2}}-1.
\end{equation}
The PHC can be obtained as
\begin{equation}
\sigma_{xz}(B)=[\sigma_{D}+\Delta\sigma(B)]\cos\theta\sin\theta.
\label{eq_sxz}%
\end{equation}
It is noted that Eqs.\ (\ref{eq_szz}) and (\ref{eq_sxz}) are applicable to the
case of strong magnetic field. For weak magnetic fields, the electrons are
weakly localized by the magnetic field, such that there could exist electric
current in the plane perpendicular to the magnetic field. However, it does not
affect our discussions on the LMC and planar Hall effect, since the electric
current in the plane perpendicular to the magnetic field mainly contributes to
the Drude conductivity $\sigma_{D}$, and, as discussed below, $\sigma_{D}$ is
almost independent on the magnetic field.

Before analyzing the numerical results, we can infer some properties of the
LMC and PHC. At low temperatures, by the approximation
$-\partial_{\epsilon}f_{0}(\epsilon)=\delta(\epsilon-E_{\mathrm{F}})$, we can
reduce $\Delta\mu$ to be
\begin{equation}
\Delta\mu=\eta^{-2}eE_{\Vert}l_{e}\frac{1}{\varTheta}%
\end{equation}
with $l_{e/a}=\upsilon_{\mathrm{F}}\tau_{\mathrm{inter}/\mathrm{intra}}$.
Subsequently, we can arrive at%
\begin{equation}
\sigma_{D}=\eta^{-2}\frac{2e^{2}}{h}\frac{eBl_{a}}{h}\Lambda(E_{\mathrm{F}})
\end{equation}
and%
\begin{equation}
\Delta\sigma(B)=\eta^{-2}\frac{2e^{2}}{h}\frac{eB(l_{e}-l_{a})}{h}%
\varTheta^{-1}. \label{eq_dsxz0}%
\end{equation}
In the weak magnetic field regime i.e., $|E_{\mathrm{F}}|\gg\hbar\omega_{c}$,
by taking the replacement $\sum\limits_{n=0}^{n_{c}}\rightarrow\int_{0}%
^{n_{c}}dn$ in Eq. (\ref{eq_vartheta}), we derive $\varTheta\simeq2(\eta
^{3}E_{\mathrm{F}}/\hbar\omega_{c})^{2}$ and $\Lambda(E_{\mathrm{F}}%
)\simeq\eta^{6}\frac{\pi hn_{e}}{eBk_{\mathrm{F}}}$, with $n_{e}%
=k_{\mathrm{F}}^{3}/3\pi^{2}$ being the carrier density. Therefore,
$\sigma_{D}=\eta^{4}\frac{e^{2}n_{e}}{\hbar k_{\mathrm{F}}}\upsilon
_{\mathrm{F}}\tau_{\mathrm{intra}}$ is just the zero-field Drude conductivity
and $\Delta\sigma(B)=\eta^{-6}\Delta\sigma_{0}(B)$ with
\begin{equation}
\Delta\sigma_{0}(B)=\frac{e^{2}}{4\pi^{2}\hbar}\frac{(eB)^{2}\upsilon
_{\mathrm{F}}^{2}}{E_{\mathrm{F}}^{2}}\upsilon_{\mathrm{F}}\tau
_{\mathrm{inter}} \label{eq_ds0}%
\end{equation}
being the LMC for $\theta=0$, where we have neglected the term tied to
$\tau_{\mathrm{intra}}$ in $\Delta\sigma_{0}(B)$ for $\tau_{\mathrm{intra}}%
\ll\tau_{\mathrm{inter}}$. Therefore, the LMC and PHC
reduce to the classical form%
\begin{align}
\Delta\sigma_{zz}(B)  &  =(1-\tanh^{2}\vartheta)^{2}\Delta\sigma_{0}%
(B)\cos^{2}\theta,\label{eq_dszz}\\
\Delta\sigma_{xz}(B)  &  =(1-\tanh^{2}\vartheta)^{2}\Delta\sigma_{0}%
(B)\cos\theta\sin\theta. \label{eq_dsxz}%
\end{align}
As can be seen, the amplitude of the LMC has $B^{2}$ dependence for any value
of $\theta$ except for $\theta=\pi/2$, and the $B$-quadratic dependence holds
for the PHC when $\theta\neq0,\pi/2$. Similar formula
were shown by Nandy $et$ $al$.\cite{PRL176804}, which predicted a $\cos
^{2}\theta$ angular dependence of the LMC and a $\cos\theta\sin\theta$ angular
dependence of the PHC. However, the
experimentally-observed angular dependence of the LMC appeared to be much
stronger than the theoretically-predicted $\cos^{2}\theta$, which is not quite consistent with the expectations drawn from the previous theory\cite{Xiong413,ncomms10735}. As different from the formula
derived by Nandy $et$ $al$., our results, in addition to the $\cos^{2}\theta$
and $\cos\theta\sin\theta$ factors, contain another angle-dependent factor
$(1-\tanh^{2}\vartheta)^{2}$. In the weak magnetic limit, e.g., $B\rightarrow
E/\upsilon_{\mathrm{F}}$, by the approximation $1-\tanh^{2}\vartheta\simeq
\cos^{2}\theta$, our results predict stronger angular dependence for the LMC
and PHC, with $\Delta\sigma_{zz}(B)\propto B^{2}\cos
^{6}\theta$ and $\Delta\sigma_{xz}(B)\propto B^{2}\cos^{5}\theta\sin\theta$.
For higher magnetic fields, i.e., $B\gg E/\upsilon_{\mathrm{F}}$
($\tanh\vartheta\rightarrow0$), the angular dependence returns to that
given by Nandy $et$ $al$., i.e., the $\cos^{6}\theta$ ($\cos^{5}\theta
\sin\theta$) angular dependence of the LMC (PHC) will
cross over to the $\cos^{2}\theta$ ($\cos\theta\sin\theta$) dependence.

The above inferences are confirmed by the numerical results shown in Fig.
\ref{figLMC}, where we plot the calculated $\Delta\sigma_{zz}(B)$ and
$\Delta\sigma_{xz}(B)$ as functions of $\theta$ for several values of the
normalized Fermi energy $E_{\mathrm{F}}/\epsilon_{0}$ and the magnetic field
$B$. As demonstrated in the classical formula, i.e., Eqs. (\ref{eq_ds0}%
)-(\ref{eq_dsxz}), the amplitude of the LMC and PHC, in
addition to the $B$-quadratic dependence, are scaled with $1/E_{\mathrm{F}%
}^{2}$, which show an unphysical diverging character if $E_{\mathrm{F}}=0$, as
presented by the red dash line in Fig. \ref{figPHE}(a). In the ultra-quantum
limit, however, the LMC and PHC remain finite for
$E_{\mathrm{F}}=0$, as can be seen from the dark-solid curves in Figs.
\ref{figLMC}(a), (c) and Fig. \ref{figPHE}(a). In fact, the case of
$E_{\mathrm{F}}=0$ is equivalent to the strong magnetic field regime, in which
only the $n=0$ LL is crossed by the Fermi level. In this limiting case, Eq.
(\ref{eq_dsxz0}) can be simplified to
\begin{equation}
\Delta\sigma(B)\simeq\eta^{-2}\frac{2e^{2}}{h}\frac{eB\upsilon_{\mathrm{F}%
}\tau_{\mathrm{inter}}}{h},
\end{equation}
which predicts an $E_{\mathrm{F}}$-independence and $B$-linear dependence of
the LMC and PHC, as also shown by Fig. \ref{figPHE}(b).
The $B$-linear dependence of the positive LMC has been observed recently,
based on Weyl orbits in \textrm{Cd}$_{3}$\textrm{As}$_{2}$\cite{Zhang:2018aa}.
For a finite but relative small $E_{\mathrm{F}}$, e.g., $E_{\mathrm{F}}%
\sim0.5\epsilon_{0}$, a step change occurs in the LMC and planar Hall
conductivity when $\theta$ reaches a critical value, as indicated by the
red-dash curves in Figs. \ref{figLMC}(a) and (c). To explain the sudden change
of $\Delta\sigma_{zz}(B)$ and $\Delta\sigma_{xz}(B)$, we plot $\Delta
\sigma(B)$, the amplitude of the LMC and PHC, as
functions of the normalized Fermi level $E_{\mathrm{F}}/\epsilon_{0}$ in Fig.
\ref{figPHE}(a). As it shows, $\Delta\sigma(B)$ oscillates strongly with
$E_{\mathrm{F}}$, due to the van Hove singularities in the DOSs. As shown by
Eq. (\ref{eq_dsxz0}), the DOSs enter $\Delta\sigma(B)$ as a denominator. As a
consequence, when the Fermi level encounters a van Hove singularity where the
DOSs increase dramatically, $\Delta\sigma(B)$ will be suppressed heavily. For
fixed $E_{\mathrm{F}}$ and $B$, with $\theta$ increasing from $0$, the van
Hove singularities, as depicted in Fig. \ref{figPHE}(a), would shift towards
the zero energy point, and then sweep over the Fermi level, leading to the
sudden drops in Figs. \ref{figLMC}(a) and (c). When $E_{\mathrm{F}}$ is in the
vicinity of the van Hove singularities, the LMC, as $\theta$ shifts away from
$0$, decays rapidly with a decaying rate far faster than $\cos^{2}\theta$, as
seen from the blue-dash-dot curve in Fig. \ref{figLMC}(a). For larger
$E_{\mathrm{F}}$, with $\theta$ or $B$ varying, more van Hove singularities
pass through the Fermi level, which results in the oscillation behavior of
$\Delta\sigma(B)$ in Fig. \ref{figPHE}(b), as also can be seen from the
blue-dash-dot curves in Fig. \ref{figLMC}. As shown by Figs. \ref{figLMC}(c)
and (d), the classical angular dependence of the LMC and planar Hall
conductivity can be recovered by increasing the magnetic field.

With further increasing the magnetic field, the number of the LBs intersected
by the Fermi level will decrease, and as a result, the magnetic-field
dependence of the LMC and PHC would, gradually, deviate
from the $B$-quadratic dependence, as demonstrated in Fig.\ \ref{figPHE}(b).
Though the LMC oscillates strongly with $B$, its classical $B$-quadratic dependence for weak magnetic field limit can be reflected by the envelope of
$\Delta\sigma(B)$, as shown in the
inset of Fig. \ref{figPHE}(b). In the strong
magnetic field regime $\hbar\omega_{c}>|E_{\mathrm{F}}/\sqrt{2}|$, where the
Fermi level only crosses the $n=0$ LL, the $B$-quadratic dependence of
$\Delta\sigma(B)$ will finally cross over to the $B$-linear dependence, as
shown by Fig. \ref{figPHE}(b).

When the magnetic and electric fields are collinear, $\Delta\sigma(B)$
exhibits a periodic-in-$1/B$ oscillation, with the period given
by\cite{DengPRL}%
\begin{equation}
\Delta(\frac{1}{B})=2e\hbar(\frac{\upsilon_{\mathrm{F}}{}}{E_{\mathrm{F}}%
})^{2}. \label{eq_B}%
\end{equation}
The periodic-in-$1/B$ oscillation is depicted by the dark solid curve of Fig.\
\ref{figPHE}(c). The periodic-in-$1/B$ oscillation is attributable to the van
Hove singularities, i.e., Eq.\ (\ref{eq_B}) is solved via Eq.\ (\ref{eq_van}) by
setting $\theta=0$.  When the magnetic and electric
fields are noncollinear,  the
classical $B$-quadratic dependence of the LMC and PHC will be
modified even in the weak magnetic regime, as shown by the cyan dash-dot
curve of Fig. \ref{figPHE}(c).

\section{summary}

\label{SUM}

In summary, based on the theory developed recently, we have studied the
properties of the magnetotransport in WSMs. It is found that the LMC and
PHC are, respectively, scaled with $B^{2}\cos^{6}\theta$
and $B^{2}\cos^{5}\theta\sin\theta$ for weak magnetic field regime. For higher
magnetic fields, the angular dependence of the LMC and PHC
cross over to $\cos^{2}\theta$ and $\cos\theta\sin\theta$
dependence. In the strong magnetic field regime, when the Fermi level is
slightly way from the Weyl nodes, a step change would occur in the LMC and
PHC, as $\theta$ reaches a critical value. With $\theta$
increasing from $0$, the $B$-quadratic dependence of the LMC and
PHC will be modified even in the weak magnetic regime.

\section{acknowledgements}

This work was supported by the National Natural Science Foundation of China
under Grants No. 11474106 (R.-Q.W), No. 11674160 (L.S.), No. 11804130 (W.L.),
No. 11804101 (W.Y.D.), the Key Program for Guangdong NSF of China under Grant
No. 2017B030311003 (R.-Q.W) and GDUPS(2017).

\bibliography{bibphe}

\bigskip

\bigskip

\end{document}